\documentstyle[aps]{revtex}
\begin{document}
\title{EFFECTS OF SHORT RANGE CORRELATIONS ON Ca ISOTOPES}
\author{G.A. Lalazissis$^{3}$, S. E. Massen$^{1,2}$\\   
\  \\
$^{1}$Nuclear Physics Laboratory, Department of Physics\\
University of Oxford, Oxford OX1-3RH, United Kingdom.\\
\ \\
$^{2}$Department of Theoretical Physics\\ 
Aristotle University of Thessaloniki, 
GR 54006 Thessaloniki, Greece\\
\ \\
$^3$Physik Department, Technische Universit\"at M\"unchen \\
D 85747 Garching, Germany}
\maketitle
\date{}
\begin{abstract}
The effect of Short Range Correlations (SRC) on Ca isotopes is studied 
using a simple phenomenological model. Theoretical expressions for
the charge (proton) form factors, densities and moments of Ca nuclei are 
derived. The role of SRC in reproducing the empirical data for the 
charge density differences is examined. Their influence on the 
depletion of the nuclear Fermi surface is studied and the fractional
occupation probabilities of the shell model orbits of Ca nuclei
are calculated. The variation of SRC as function of the mass
number is also discussed.  
\end{abstract}
\newpage 
\section{Introduction}
Calcium nuclei have been of great experimental as well as theoretical 
interest. It is the only magic element for which precision measurements 
on isotope shifts \cite{o89,b95} (see also \cite{Andl82})  have been 
carried out over a full neutron shell, namely the 1f$_{7/2}$ shell between
the two doubly magic isotopes $^{40}$Ca and $^{48}$Ca. 

The empirical data for the isotope shifts \cite{o89} show an anomalous 
A dependence. The addition of neutrons to the $^{40}$Ca core leads to 
an increase of the charge radii up to $^{44}$Ca. Then adding more neutrons 
the charge radii start to decrease. The very interesting feature is that 
the charge radii of the two doubly magic nuclei  $^{40}$Ca and $^{48}$Ca 
have  practically the same value. It is noted, however, that the electron 
scattering experiments have shown that the charge distributions 
of these magic nuclei are not identical \cite{Em83}. Moreover, Ca isotopes 
demonstrate clearly the very interesting differential effect of the 
odd-even staggering of nuclear radii. That is, the mean square (MS) 
radii of the odd neutron nuclei are smaller than the average of their 
even neutron neighbors. 

Apart from laser spectroscopy, which provides very accurate experimental  
information about the isotope shifts of Ca nuclei, other experimental 
techniques (muon spectroscopy, electron and hadron scattering) offer
additional information on the charge and mass distributions \cite{re79}.   
Empirical  data for the form factors and their isotopic change for some 
even stable isotopes is available. Therefore, the rich experimental input 
makes Ca nuclei attractive for theoretical study. 

Brown et al. \cite{Brown79} calculated the charge
distribution of the Ca isotopes using a Woods-Saxon state dependent
potential with a density dependent symmetry potential which was determined
in a self consistent way and using non integer occupation probabilities for
the $1d_{3/2}$, $1f_{7/2}$, and $2p_{3/2}$ states. Bhattacharya et al. \cite
{Bhattacharya93} using an average one-body potential of Woods-Saxon type and
experimental occupation probabilities, have reproduced the parabolic
variation of the charge radii of the Ca isotopes.
Zamick \cite{Zamick71} and Talmi \cite{Talmi84}, in analogy with
the binding energies, assumed that the effective radius operator
has a two body part as well as one-body part. They were able to
explain the odd-even staggering effect observed in Ca isotopes
assuming that a mechanism which gives rise to this odd-even
variation is the  polarization of the core by the valence neutrons.
Finally Barranco and Broglia \cite{Barranco85}, in perhaps the
most fundamental approach, succeeded to explain the
parabolic variation of the MS charge radii introducing
collective zero-point  motion.

Mean field calculations fail to reproduce the parabolic behavior of the 
charge radii of the Ca isotopes. This is an indication that one
has to go  beyond the Hartree-Fock approach taking into account nuclear 
correlations. There are various types of correlations. In Refs.
\cite{Zamick71,Talmi84,Barranco85} ground state correlations
have been accounted and their effect to the reproduction of the
empirical data for the isotope shifts of Ca nuclei was
investigated. However, other types of nuclear correlations might
also be important. Such as, short range correlations (SRC) which
describe the effect of the distortion of the relative
two-body wave function at small distances.

Correlated charge form 
factors, $F_{ch}(q)$, and densities of s-p and s-d shell nuclei were
calculated \cite{SEM88,SEM89,SEM90,SEM90b,LMP94}, using  Jastrow type
correlations \cite{Jast55} for the correlated wave
functions of the relative motion and employing the
factor cluster expansion of Ristig, Ter Low, and Clark 
\cite{Ristig-Clark,Clark79}.
First the method  was applied to the doubly closed shell light
nuclei $^{4}$He , $^{16}$O and $^{40}$Ca and then it was
extended in an approximate way to all nuclei in the region
4 $\leq$ A $\leq$ 40.  

In the present work the method is expanded to the study of the
isotopes of the closed shell nuclei.
The isotopic chain of Ca nuclei has been chosen 
for the study,  due to its special interest.
The calculated values of the differences of the charge density
distributions of even Ca nuclei are compared with the available
experimental data \cite{Frosch68,Em83}.
Our aim is to examine  the effect of SRC on Ca nuclei
as well as their variation with the mass number.

High resolution electron scattering experiments have shown deviations from
the mean field picture \cite{f83p84w90,f91}. The quantum states,
especially those near the Fermi surface, appear to be depleted. This is a 
clear demonstration that the single-particle orbits are partially occupied 
because of the nucleon-nucleon correlations \cite{f91}. The depletion 
of the occupied states can be  attributed to a coupling of the 
~Hartree-Fock ground state to low-lying collective modes and to
SRC due to hard  collisions between nucleons at small distances.
Calculations for nuclear matter including SRC have shown \cite{Ra89,Fa89} 
that the depletion of the otherwise filled orbits is 10-20 \%.

Recently, a simple method has been proposed  for  the
determination of the fractional occupation probabilities of the shell model 
orbits of the ground state wave function as well as the total
depletion of the nuclear Fermi sea \cite{LMP92,LMP93} (see also \cite{LMP94}).
The correlated densities are used as an input and the connection
is done by means of the ``natural
orbital'' representation \cite{low55}, which allows us to keep the
simplicity of the shell model picture.
Here, employing this method, we  determine the occupation
numbers of the shell model orbits and the total depletion of
the nuclear Fermi sea. The latter reflects the influence of SRC
 to the deviation from the mean field approach. The variation 
of these quantities with the mass number is also studied.    

The paper is organized as follows. 
In Sec. 2 the relevant formalism of SRC for the closed shell
nuclei is shown. 
In Sec. 3 the expansion to the isotopes of the closed shell
nuclei is presented. In Sec. 4 the method for the determination
of the occupation numbers is briefly discussed. 
 In Sec. 5 the numerical results are reported and discussed.
Finally, Sec. 6 summarizes our main  conclusions.


\section{ Correlated charge form factor, densities and MS radii
for closed shell nuclei}

Expression for the correlated charge form factors, $F_{ch}(q)$, 
of the closed s-p and s-d shell  nuclei
were derived \cite{SEM88,SEM89,SEM90,SEM90b,LMP94} in the
 framework of the factor
cluster expansion
of Ristig, Ter Low and Clark \cite{Ristig-Clark,Clark79} using
the Jastrow ansatz for the correlated wave-functions. 
This type of correlations is characterized
by the correlation parameter $\lambda_{nlS}$ which enters in the
normalized correlated wave functions of the relative motion: 
\begin{equation}
\label{Jastrow wf}\psi _{nlS}(r)=N_{nlS}[1-\exp (-\lambda_{nlS} 
r^2/b^2)]\phi_{nl}(r) 
\end{equation}
where $N_{nlS}$ are the normalization factors, $\phi _{nl}(r)$
are the harmonic oscillator (HO) wave functions and  $b=\sqrt{2}b_1$ 
($b_1=\sqrt{\hbar /m\omega }$) is the harmonic oscillator
parameter for the relative motion.  
In this approach the expression for the point proton form factor, $F(q)$,  
takes the form: 
\begin{equation}
\label{f-src}F(q)=F_1(q)+F_2(q) 
\end{equation}
where 
\begin{equation}
 F_{1}(q)= {1\over Z } \exp{[-{b_{1}^{2} q^{2} \over 4 }] } 
 \sum_{k=0}^{2} N_{2k} ({ {b_{1} q} \over 2 })^{2k}    
\label{f-HO}
\end{equation}
is  the contribution of the one-body term to $F(q)$ with
\begin{equation}
\begin{array}{lll}
N_{0}=2(\eta _{1s} + \eta _{2s} + 3 \eta _{1p} + 5 \eta _{1d}) & , &
N_{2}=-{4 \over 3} (2 \eta _{2s} + 3 \eta _{1p} + 10 \eta _{1d})  \\
 & &  \\
 N_{4}= {1 \over 3} (4 \eta _{2s} +8 \eta _{1d}) & &
\end{array}
\label{f-HO-N}
\end{equation}
and $\eta_{nl}$ is the occupation probability (0 or 1 in the present case) of 
the $nl$ state.

The contribution of the two-body term, $F_2(q)$, to the form factor  
$F(q)$ can be expressed in a rather simple way in closed form by
means of the matrix elements: 
$$
A_{nlS}^{{n}^{\prime }{l}^{\prime }{S}^{\prime }}(j_k)=<\psi
_{nlS}|j_k(qr/2)|\psi _{{n}^{\prime }{l}^{\prime }{S^{\prime }}}> 
$$
These are simple polynomials and exponential functions of $q^2$ \cite
{SEM88,SEM89,SEM90,SEM90b,LMP94}.
In our study the correlation parameter $\lambda_{nlS}$ is taken 
state independent ($\lambda_{nlS}$ = $\lambda$). It is noted
that  it has been  shown in \cite{LMP94}, that the effect of the 
state dependence of the short range correlations is small.

Then the charge form factor, $F_{ch}(q)$, is written :
\begin{equation}
\label{chff} F_{ch}(q)= f_p(q) \times f_{CM}(q) \times F(q) 
\end{equation}
where $f_p(q)$ and $f_{CM}(q)$ are the corrections 
due to the finite proton size \cite{SEM88} and the centre of
mass motion \cite{Tassie58} respectively.
The correlation parameter $\lambda $ and the HO parameter $b_1$ are
determined by fitting formula (\ref {chff}) to the experimental
data of $F_{ch}(q)$.

An interesting feature of this method is the possibility of
finding an analytic form for the correction to the uncorrelated
charge (proton) density distribution by means of a Fourier
transform of $F_{2}(q)$. Thus the correlated proton density
distribution is written:

\begin{equation}
\label{den-src}\rho_{cor} (r)=\rho _1(r)+\rho _2(r) 
\end{equation}
Analytic expressions can also be derived for the various
moments of the charge (proton) density distribution. 
The moments have the form:
\begin{equation}
\label{r-src}<r^k>=<r^k>_1+<r^k>_2 
\end{equation}
where $<r^k>_1$ and $<r^k>_2$ are the contributions of the one
and two-body density  $\rho _1(r)$ and $\rho _2(r)$  respectively.

A very satisfactory approximate
expression for the MS charge radius is also derived  
\begin{equation}
\label{r2aprox-1}<r^2>_{ch}= C_{HO}(1-\frac
1A)b_1^2+C_{SRC}b_1^2\lambda ^{-3/2}+r_p^2+\frac NZr_n^2
\end{equation}
where  $r_p^2$ and $r_n^2$ are the proton and neutron MS
charge radii respectively. For the latter the value 
$r_n^2$ = $-$0.116 fm$^{2}$ is used  \cite{Chandra76}.
  The constants $C_{HO}$ and  $C_{SRC}$
for  $^{40}$Ca are: $C_{HO}$= 3 and $C_{SRC}$= 12.4673.
It is noted that the values obtained with formula
(\ref{r2aprox-1}) differ less than 0.08\% from those obtained
with the exact expression for the radii and therefore formula
(\ref {r2aprox-1}) is suitable for practical use.

The merit of the approach is mainly the simplicity, as most of
the calculations are analytic. Another advantage is the
possibility of obtaining 
approximate expressions of the two body term of various
quantities by  expanding the expression of the form factor in powers of 
$\lambda $. This allows the study of the open shell nuclei in
the region 4 $\leq$ A $\leq$ 40 \cite{SEM90,SEM90b}.

\section {Extension to the isotopes of closed shell nuclei}

The method described in the previous section is exactly
applicable to doubly closed s-p and s-d shell nuclei and
approximately to open shell nuclei in the region
4 $\leq$ A $\leq$ 40.   
Here, we extend the method to the study of the
isotopes of the closed shell nuclei. 

In an isotopic chain all nuclei have the same atomic number Z. Thus,
we assume that the correlated 
charge (proton) form  factors, densities and moments of the isotopes can be 
described by the same formulas (\ref{f-src}), (\ref{den-src}), and (\ref
{r-src}) where, however, different values for the parameters
$\lambda$ and $b_{1}$ are used. 
The correlation parameters as well as the size parameters for the
isotopes  could be easily determined if empirical data for the charge form
factors (especially for high momentum transfers) of all the
isotopes  were available. However, usually this is not
the case and therefore one has to try other possibilities.

First, the correlation parameter is written in a more convenient way: 
\begin{equation}
\label{mu1} \mu =\sqrt{\frac{b_1^2}\lambda } 
\end{equation}

\noindent Then we assume that the correlation parameters $\mu (A_c+n)$ and 
the HO parameters $b_{1} (A_c+n)$ of the isotopes can be  written:

\begin{equation}
\label {mu} \mu (A_c+n) = \mu (A_c) + \delta \mu (A_c+n) \\ 
\end{equation}
\begin{equation} 
\label {b1} b_{1} (A_c+n) = b_{1} (A_c) + \delta b_{1} (A_c+n)  
\end{equation}
\noindent
where $\mu (A_c)$ and $b_{1} (A_c)$ are the parameters of the
corresponding closed shell nucleus ($A_c$).
The differences  $\delta \mu$
 and $ \delta b_{1}$ express the change in the
~parameters due to the addition of extra neutrons ($n$). 

Using (\ref{mu}) and (\ref{b1}) expression (\ref{r2aprox-1})
for the MS charge radius,  $<r^{2}>_{ch} = r^{2}$, can 
be  written in the following way: 
\begin{eqnarray}
\label{r2}r^2(A_c+n) &= & r^{2} (A_{c}) + \delta r^{2} (A_{c}+n)
\nonumber\\
&  = & C_{HO} (1- \frac{1}{A_c+n})
[b_1(A_c) + \delta b_1(A_c+n) ]^2 +  \nonumber\\
& &C_{SRC} \frac{[\mu(A_c) +\delta \mu (A_c+n) ]^3} {b_1(A_c) + 
\delta b_1 (A_c+n)}   + r^2_p + \frac{A_c - Z +n}{Z} r^2 _n
\label{r2aprox-3}
\end{eqnarray}

For the closed shell nuclei experimental data is available in
most of the cases. The parameters $\mu (A_{c})$ and
$b_{1} (A_{c})$ can be determined by a fit to the data. 
Thus the problem is  reduced to the determination of the differences  
$\delta b_1$ and $\delta \mu $. For this, additional input is
necessary. In the present approach the differences are determined 
using the empirical data for the isotope shifts and isospin dependent 
theoretical expressions for the oscillator parameters. 

In a very recent publication \cite{LP95} new improved expressions for the
oscillator spacing $ \hbar \omega$ were derived. These expressions have the
advantage of being isospin dependent. They were obtained by
employing new expressions for the MS radii of nuclei, which
fit the experimental MS radii and the isotope shifts much
better than other frequently used relations.   
In the present work the following formula for $\hbar\omega$
taken from \cite{LP95} is used:

\begin{equation}
\label{hbar-omega}\hbar \omega = 38.6 A^{-1/3} \left [ 1 + 1.646 A^{-1} -
0.191 (N-Z) A^{-1} \right ] ^{-2} 
\end{equation}
The derivation of the HO parameters of the isotopes by means
of (\ref{hbar-omega}) is straightforward: 

\begin{equation}
\label{b1-n1}b_1(A_c+n) = \sqrt{\frac{\hbar ^2}{38.6 m} } A_c^{1/6} (1+\frac{%
n}{A_c})^{1/6}  \left[ 1 + \frac{1.646}{A_c} (1 + \frac{n}{A_c} )^{-1} -
0.191 \frac{n}{A_c}  ( 1 + \frac{n}{A_c} )^{-1} \right ] 
\end{equation}

Formula (\ref{b1-n1}) is used for the determination of the differences 
$ \delta b_{1} (A_c+n)$. Finally, the isotopic changes $ \delta \mu
(A_c+n)$  of the correlation parameters  are
adjusted to reproduce the empirical data of the isotope shifts,
$\delta r^{2} (A_{c} + n)$, 
using equation (\ref{r2}).   
Hence, in the present approach, the parameters $b_{1} (A_{c}
+n)$ and $\mu (A_{c} + n)$, calculated from (\ref{mu}) and
(\ref{b1}) respectively, are determined using the empirical
data for the charge radii of the isotopic chain and the
information obtained from the experimental charge form factor of
the closed shell isotope. 

\section {The method for the determination of the occupation numbers} 

The correlated proton densities $\rho_{cor}(r)$ can be used as an input for 
the determination of the fractional occupation probabilities of the shell 
model orbits of the ground state wave function \cite{LMP92}. The connection 
with short range correlations is done by employing the ``natural orbital''
representation \cite{low55}. The natural orbital approach has already been
applied for nuclear studies in the past \cite{gau71} (see also \cite{Sto93}).
Recently it was also employed \cite{pan88} within a ~variational Jastrow-type
correlation method for the study of quantum liquid drops.

For spherical symmetric systems the density distribution in the natural 
orbital representation $\rho_{n.o.}(r)$  takes the  simple form 
\begin{equation}
\rho_{n.o.}(r) = {1 \over 4\pi}\sum_{nl}(2j+1)\eta_{q}|\phi_{q}(r)|^{2},
\end{equation}
where $\eta_{q}$ is the occupation probability ($\eta_{q}$ $\leq$ 1) of the 
$q (=nlj)$ state.
The ``natural orbitals'' $\phi_{q}$ are approximated by the radial part of 
the single-particle wave functions \{R$_{nl}$\} of a harmonic oscillator 
potential. The occupation probabilities are determined by  assuming  
$\rho_{cor}(r) = \rho_{n.o.}(r)$. That is, the correlated proton density
distribution, $\rho_{cor}(r)$, in which the effect of short range
correlations is taken into account, equals with density distribution 
$\rho_{n.o.}(r)$, corresponding to the natural orbital representation. We 
demand the first few moments of $\rho_{cor}(r)$ to be equal to those of
$\rho_{n.o.}(r)$ distribution. 
\begin{equation}
<r^{k}>_{cor} = <r^{k}>_{n.o.}  
\end{equation}
The details of the calculations are described in Refs.
\cite{LMP92,LMP93}. 
The merit of this approach is that by ``mapping'' the correlated density 
distributions to those calculated with the ``natural orbitals'' a 
relationship is established between the fractional occupation probabilities 
and the short range correlations. The effect of short range correlations is 
taken into account in an effective way and it is absorbed in the values of 
the calculated occupation numbers and the size parameters b$_{n.o.}$. It is 
a suitable way of keeping the simplicity and visuality of the 
single-particle picture. It should be noted, however, that this 
relationship is not completely clear because one is not able to distinguish 
the correction to the charge form factor $F_{ch}(q)$ for large $q$
because of short range correlations from the one due to meson ~exchange 
currents. In addition, the use of harmonic oscillator wave functions
is a simplification. Proper linear compinations of oscillator wave functions 
could be used instead. In such a case, however, the method loses its
simplicity.

The method was improved \cite{LMP93} by considering different oscillator
parameters for the hole states and those above the Fermi sea
(FS).  Specifically
the $\rho_{n.o.}(r)$ was divided in two parts: 

\begin{equation}
\rho_{n.o.}(r) = \rho_{n.o.}^{<FS}(r) + \rho_{n.o.}^{>FS}(r)
\end{equation}
where each part is expressed by a harmonic oscillator basis characterized by
the oscillator parameters b and \~b respectively.
In addition, it was assumed that for $\rho_{n.o.}^{<FS}(r)$ the occupancy of
the states above the Fermi level is practically zero, while for
$\rho_{n.o.}^{>FS}(r)$ only the states above the Fermi sea have occupancies
which appreciably differ from zero. The two parts of
$\rho_{n.o.}(r)$ in (17),  
somehow, reflect nuclear characteristics which are sensitive to the low and
high momentum component of the charge form factor respectively. 
In the present work the formalism of Ref. \cite{LMP93} has been adopted.


\section{Numerical results and discussion}

In table I the correlation parameters $\mu (A_{c}+n)$  and the HO
parameters $b_{1} (A_{c}+n)$ for all Ca isotopes considered in
this approach are shown. Their calculation has been done with
the aid of formulas (\ref{b1}) and (\ref{mu}).
The parameters $\mu (A_{c})$ and $b_{1} (A_{c})$ of the closed
shell nucleus (core nucleus $A_{c}$ = 40) were determined by
fitting the theoretical expression of the charge form factor,
$F_{ch}(q)$, of $^{40}$Ca to the experimental data. The
differences $\delta b_{1}$ are determined using formula   
(\ref{b1-n1}) while  
$\delta \mu $ are adjusted so that expression (\ref{r2}) for the
MS charge radii to reproduce the empirical data
of the isotope shifts, $\delta r^{2}$, of Ca nuclei \cite{o89}.

In Fig. 1 the isotopic change $\delta \mu$ of the correlation 
parameter $\mu$ as function of the mass number is shown. It is observed
that the parameter $\mu$, which expresses the strength of the
short range correlations, is increasing up to $^{44}$Ca and then starts
decreasing following the same variation with the
isotope shifts (see for example Fig 49 of \cite{o89}).
This indicates that there is a proportion between the ~strength of SRC and the
size of the nucleus, that is when SRC become stronger the charge
radii become larger. It is interesting to note that, if for the
determination of $b_{1}$ isospin independent
expressions of $\hbar\omega$ (see for example Refs. \cite{mol68,das83,lp93})
were used instead, the two quantities do not show the same
variation. In this case, as the mass number increases the
correlation parameters have opposite variation compared with the one
of the charge radii. On the other hand the use of such
expressions for $ \hbar \omega$ does not provide a satisfactory description 
of the empirical data for the charge density difference of 
$^{48}$Ca $-$ $^{40}$Ca.            
 
Using the values of Table I the charge (proton) form  factors,
density distributions as well as the differences of the density
distributions $\Delta \rho$ (40+n) = ($\rho$ (40+n)$-$ $\rho$ (40))
can be easily calculated. In Figs. 2-3 the quantity  
$\Delta \rho_{ch} (40+n)r^2$ for the charge distribution differences of 
$^{42}$Ca $-$ $^{40}$Ca and  $^{44}$Ca $-$ $^{40}$Ca ~respectively are 
compared with the empirical data (solid lines). The same is also in Fig. 4 
for the difference $^{48}$Ca $-$ $^{40}$Ca. In this case 
the available experimental values correspond to the proton density
distributions. The two solid lines correspond to the upper and
lower values of the proton density difference.  It is seen that the 
theoretical curves (dashed lines)  show the correct trend.  The calculated  
$\Delta \rho _{ch}(40+n)r^2$ reproduce the behavior of the data. That is, 
the charge flows from the center (and the outer skin in $^{48}Ca$) into a 
region around the half-density radius.  The comparison is not very
good in all cases. Especially in Fig. 4 for the difference
$^{48}$Ca $-$ $^{40}$Ca, where the maximum is not reproduced
well. However, in the present approach SRC are only accounted.
This indicates that additional correlations, as for example surface
vibrations, are necessary to improve the agreement with the experiment. 
In this analysis only a  small part of their effect, ``hidden'' in the 
~empirical data of the isotope shifts (used for the determination of 
$\delta \mu$), might be accounted. It should be noted, however, that in 
this approach the parameters $\mu$ and $b_{1}$ are not free. If one or both 
of them were free, then one could expect better agreement. In such
a case the effect of other type correlations would have been taken into 
account effectively. Thus, for example, the parameters $\mu$ and $b_{1}$ 
could be determined by direct fits to the experimental charge form factors. 
This procedure was not followed because experimental data for the charge 
form factors of all Ca nuclei is not available. Moreover, this data
do not cover the region of the high momentum transfers (q $>$ 3fm$^{-1}$). 

Next, the calculated proton density distributions
$\rho_{cor}(r)$ are used for the determination of the occupation
numbers of the shell model orbits of Ca nuclei following the
procedure described in Refs. \cite{LMP92,LMP93}.  
In table II the calculated size parameters b, and \~b of the two
natural orbital bases are shown together with the occupation ~probabilities  
of Ca nuclei.  The total depletion of the nuclear Fermi sea, which reflects 
the effect of the short range correlations and gives a measure of the 
deviation from the mean field picture is also shown. 

In Fig. 5 the relative depletion of Ca isotopes  is plotted against the 
mass number A. As relative depletion we define the quantity:

 relative depletion = [ depl. ($^{40+n}$Ca) $-$ depl.
($^{40}$Ca) ] / depl.($^{40}$Ca).

\noindent A clear parabolic shape analogous to
that of Fig. 1 is observed. This is because the
depletion of the nuclear Fermi sea expresses the effect of SRC
and thus a similar variation with the correlation parameters
$\mu$  should be expected.  

In Fig. 6 the variation with the mass number of the occupation
probabilities of the shell model orbits of Ca nuclei is shown. 
For the very deep states the ~occupation
probabilities ( see also Table II)
are very close to one while the surface levels
deviate significantly from unity, manifesting thus the effect of SRC.   
Their variation with the mass number shows the correct behavior
and it is consistent with Fig. 1 for the correlation parameters
$\mu$, which measures the strength of SRC. That is for
stronger SRC a larger ``fraction'' of protons is moved above
the Fermi level.   

It is noted, that the occupation probabilities  are not 
directly measured quantities. The experimental occupation
probabilities are usually obtained by extrapolating the experimental
spectral functions by means of a Gaussian fit. In general they  are
of limited accuracy and the comparison with the theory varies
in the various models. The ``experimental'' information about  the 
occupation probabilities of Ca isotopes is limited to the magic
nucleus $^{40}$Ca. On the other hand, there are no theoretical
predictions in the literature (to our knowledge) for the other Ca nuclei. 
In Table III the occupation probabilities of $^{40}$Ca, calculated in this 
work, together with the ``experimental'' values \cite{kra90,mou76} and those 
from other theoretical analyses \cite{lip85,lp91,neck92} are shown for the
sake of comparison.     

Finally we note that in Figs. 1, 5 and 6 a kind of odd-even effect is also
observed. The effect of SRC appears to be
weaker for the odd nuclei compared with their even partners.
It should be noted, however, that in the framework of this
simple approach, one cannot draw easily conclusions about the
odd isotopes, where additional effects have to be taken into account.   
One could say that the correlation parameters $\delta \mu$ are adjusted
to reproduce the isotopic changes of the charge radii and
therefore are somewhat ``forced'' to follow such a variation.

\section{Summary}

In the framework of a simple phenomelogical model theoretical
expressions for the correlated charge (proton) form factors,
densities and moments of the isotopes of closed shell nuclei are
derived. SRC are accounted using Jastrow
type wave-functions for the correlated wave functions of the
relative motion. In the present work the isotopic chain of Ca
nuclei is studied and the influence of SRC on Ca isotopes is
examined by comparing with the available empirical data for the
charge (proton) density differences. The calculated values for
the differences of the density distributions show the correct trend. 
However, the present study indicates that additional
correlations could improve the description of the experimental data.

The role of SRC      
on the depletion of the Fermi sea as well as its variation with
the mass number is discussed. The
occupation probabilities of the shell model orbits of Ca nuclei
are calculated. One should keep in mind, however, the large
uncertainties concerning their experimental determination and
the model dependence of the various theoretical analyses.  
 
Concluding we would like to mention that the main 
advantage of the present analysis is its simplicity. The
method can be applied to other isotopic chains to provide
predictions for the charge form factors, charge
density differences and other quantities for which the effect of
SRC is important.

\bigskip
\bigskip
\thanks{
One of the author (SEM) would like to thank the Nuclear Physics
Laboratory  of the University of Oxford for the kind hospitality and the
Aristotle University of ~Thessaloniki for granting his sabbatical leave.
Finally, the authors thank Dr. P.E. Hodgson and Prof. M.E.
Grypeos for useful discussions.} 

\newpage

\noindent\begin{table}
\begin{center}
\caption 
{\sf The values of the HO parameters $b_1$ (in fm) and the SRC parameters 
$\mu$ (in fm) for the $Ca$ nuclei.}
\bigskip
\begin{tabular}{ll c c c ll }
\hline\hline
& A & $b_1$ & $\mu$ &\\
\hline
&40 &1.860  &0.499  &\\
&41 &1.857  &0.504  &\\
&42 &1.855  &0.548  &\\
&43 &1.853  &0.534  &\\
&44 &1.851  &0.566  &\\
&45 &1.849  &0.543  &\\
&46 &1.848  &0.545  &\\
&47 &1.846  &0.527  &\\
&48 &1.845  &0.528  &\\
\hline\hline
\end{tabular}
\end{center}
\end{table}
\noindent\begin{table}
\begin{center}
\caption{\sf The calculated harmonic oscillator parameters, b and \~b
(in fm) together with the occupation probabilities of the shell model
orbits of Ca  nuclei}
\bigskip
\begin{tabular}{l c c c c c c c c c c l}
\hline\hline
&  A      & 40   & 41  &42   &43   &44  &45   &46    &47   &48  &\\
\hline\\
&b        & 1.893&1.892&1.892&1.887&1.889&1.884&1.883&1.882&1.880&\\
&\~b      & 1.734&1.732&1.729&1.726&1.724&1.723&1.721&1.720&1.719&\\
\hline\\
&1s& 1.000 & 0.999&  0.860 & 0.983&  0.709 & 0.911 & 0.884 & 0.999&1.000&\\
&1p& 0.822 & 0.815&  0.677 & 0.643&  0.625 & 0.639 & 0.677 & 0.682&0.657&\\
&1d& 0.565 & 0.550&  0.592 & 0.643&  0.625 & 0.638 & 0.636 & 0.606&0.631&\\
&2s& 0.565 & 0.550&  0.480 & 0.448&  0.486 & 0.458 & 0.462 & 0.458&0.450&\\
&1f& 0.424 & 0.438&  0.479 & 0.448&  0.486 & 0.458 & 0.462 & 0.458&0.449&\\
&2p& 0.058 & 0.059&  0.105 & 0.097&  0.134 & 0.107 & 0.111 & 0.086&0.092&\\
\hline\\
&dip. \% & 31.45 & 32.56  & 36.72&34.24 & 38.05 & 35.26 & 35.63&34.68&34.23&\\
\hline\hline
\end{tabular}
\end{center}
\end{table}
\newpage
\noindent\begin{table}
\begin{center}
\caption{\sf Comparison of the occupation probabilities
of $^{40}$Ca calculated in this work ($\eta_{q}$)
together with ``experimental'' and ~theoretical values from
other studies}
\bigskip
\begin{tabular}{l c c c c c c c c c c l}
\hline\hline
& nl &$\eta_{q}$ & ~expt.$^{[36]}$  & ~expt.$^{[37]}$ & [38,39]   &[40]  &\\
\hline\\
&1s& 1.000 & 0.820&  0.750 & 0.970&  0.990 &\\
&1p& 0.822 & 0.767&  0.950 & 0.975&  0.986 &\\
&1d& 0.565 & 0.720&  0.770 & 0.884&  0.962 &\\
&2s& 0.565 & 0.740&  0.650 & 0.870&  0.960 &\\
&1f& 0.424 & 0.307&    -   & 0.071&  0.030 &\\
&2p& 0.058 & 0.100&    -   & 0.035&  0.010 &\\ 
\hline\hline
\end{tabular}
\end{center}
\end{table}

\newpage

\bigskip

\begin{center}
{\bf Figure Captions}
\end{center}

\bigskip
{\bf Fig. 1.} The isotopic change $\delta \mu$ of the
correlation parameter $\mu$  as function of the mass number A.

\bigskip
{\bf Fig. 2.} The difference of the charge distributions of $^{42}$Ca $-$ 
${}^{40}$Ca, multiplied by $r^2$, (dashed line) calculated in the
present approach together with the experimental data (solid
line) taken from ref. \cite{Frosch68}. 

\bigskip

{\bf Fig. 3.} The same as in Fig.2, for the charge distribution
difference  of $^{44}$Ca $-$ ${}^{42}$Ca. 

\bigskip

{\bf Fig. 4.} The difference of the point proton distributions
of $^{48}$Ca $-$ $^{40}$Ca multiplied by $r^2$ (dashed line) calculated in the
present approach together with the empirical data taken from ref. \cite
{Em83}. The two solid lines correspond to the upper and
lower values of the experimental proton density difference.

\bigskip
{\bf Fig. 5.} The variation with the mass number A of the
relative depletion of Ca nuclei. 

\bigskip
{\bf Fig. 6.} The variation with the mass number A of the
occupation probabilities of the shell model orbits of Ca nuclei. 


\newpage 

\begin{thebibliography}{99}
\bibitem{o89}E.W. Otten, in {\it Treatise on Heavy-Ion Science}, 
edited by D.A. Bromley (Plenum, New York, 1989) Vol 7, p. 515 

\bibitem{b95}J. Billowes 
and P. Campbell, J. Phys. G: Nucl. Part. Phys. {\bf 21} (1995) 707. 

\bibitem{Andl82}  A. Andl, K. Bekk, S. G\"{o}ring, A. Hanser, G. Nowicki, H.
Rebel, G. Schatz, and R.C. Thompson, Phys. Rev. C {\bf 26} (1982) 2194.

\bibitem{Em83} H.J. Emrich, G. Fricke, G. Mallot, H. Miska, and
H.G. Sieberling, Nucl. Phys. {\bf A396}, 401c (1983).

\bibitem{re79} H. Rebel, H.J. Gils, and G. Schatz (eds), Proc.
Int. Discussion-Meeting on {\it What do we know about the radial
shape of nuclei in the Ca region?} KFK-Rep. 2830, Karlsruhe (1979).

\bibitem{Brown79}  B.A. Brown, S.E. Massen, and P.E. Hodgson, J. Phys. G:
Nucl. Phys. {\bf 5} (1979) 1655.

\bibitem{Bhattacharya93}  R. Bhattacharya, and K. Krishan, Phys. Rev. 
C {\bf 48} (1993) 577.

\bibitem{Zamick71}  L. Zamick, Ann. Phys. {\bf 66} (1971) 784.

\bibitem{Talmi84}  I. Talmi, Nucl. Phys. {\bf A423} (1984) 189.

\bibitem{Barranco85}  F. Barranco and R.A. Broglia, Phys. Let. {\bf 151B} 
(1985) 90.

\bibitem{SEM88}  S.E. Massen, H.P. Nassena and C.P. Panos, J. Phys. G: Nucl.
Phys. {\bf 14} (1988) 753.

\bibitem{SEM89}  S.E. Massen, and C.P. Panos, J. Phys. G: Nucl.
Phys. {\bf 15} (1989) 311.


\bibitem{SEM90}  S.E. Massen, J. Phys. G: Nucl. Part. Phys., {\bf 16} 
(1990) 1713.

\bibitem{SEM90b} S.E. Massen, in ``First Hellenic Symposium on Theoretical
Nuclear Physics'', Thessaloniki, Greece, (1990) p.62.

\bibitem{LMP94} G.A. Lalazissis, S.E. Massen, and C.P. Panos,
Z. Phys. A  {\bf 348}, 257 (1994) 

\bibitem{Jast55} R. Jastrow, Phys. Rev. {\bf 98}, (1995) 1479.

\bibitem{Ristig-Clark}  M.L. Ristig, W.J. Ter Low and J.W. Clark, Phys. Rev. 
C {\bf 3} (1971) 1504.

\bibitem{Clark79}  J.W. Clark, Prog. Part. Nucl. Phys. {\bf 2} (1979) 89.


\bibitem{Frosch68}  R.F. Frosch, R. Hofstadter, J.S. McCarthy, G. K.
N\"{o}ldeke, K.J. Van Oostrum, M.R. Yearian, B.C. Clark, R. Herman, and 
D.G. Ravenhall, Phys. Rev. {\bf 174} (1968) 1380 .


\bibitem{f83p84w90} B. Frois, J.M. Cavedon, D. Goutte, M. Huet, P. Leconte,
C.N. Papanicolas, X.H. Phan, S.K. Platchkov, and S.E. Williamson,
Nucl. Phys. {\bf A396}, 409c (1983); V. R. Pandharipande, C.N. Papanicolas, and
J. Wambach, Phys. Rev. Lett. {\bf 53}, 1133 (1984); P.K.A. de Witt Huberts,
J. Phys. G: Nucl. and Part. Phys. {\bf 16}, 507 (1990).

\bibitem{f91} B. Frois, Nucl. Phys. {\bf A522}, 167c (1991).

\bibitem{Ra89}  A. Ramos, A. Polls, and W.H. Dickhoff, Nucl. Phys. 
{\bf A503} (1989) 1.

\bibitem{Fa89}  A. Fabrocini and S.Fantoni, Nucl. Phys. {\bf A503}
(1989) 357.
\bibitem{LMP92} G.A. Lalazissis, S.E. Massen, and C.P. Panos,
Phys. Rev. C {\bf 46}, 201 (1992).

\bibitem{LMP93} G.A. Lalazissis, S.E. Massen, and C.P. Panos,
Phys. Rev. C {\bf 48}, 944 (1993).

\bibitem{low55} P.-O. L\"owdin, Phys. Rev. {\bf 97}, 375 (1955).

\bibitem{Tassie58}  L. J. Tassie and F. C. Barker, Phys. Rev {\bf 111} 
(1958) 940.

\bibitem{Chandra76} H.Chandra, and G. Sauer, Phys. Rev. C {\bf
13} (1976)  245.

\bibitem{LP95}  G.A. Lalazissis and C.P. Panos, Phys. Rev. C  {\bf 51}
 (1995)  1247.

\bibitem{gau71} M. Gaudin, J Gillespie, G Ripka, Nucl. Phys. {\bf 176}, 237
(1971); F Malaguti, A. Uguzzoni, E. Verondini, P.E Hodgson, Nuovo Cimento {\bf
5}, 1 (1982).

\bibitem{Sto93}  M.V. Stoitsov, A.N. Antonov, and S.S. Dimitrova, Phys.
Rev. C {\bf 48} (1993) 74.

\bibitem{pan88} D.S Lewart, V.R. Pandharipande, S.C. Peiper, Phys. Rev. B {\bf
37}, 4950 (1988).

\bibitem{mol68}  J. Blomqvist and A. Molinari, Nucl. Phys. {\bf A106}
(1968) 545 .

\bibitem{das83}  C.B. Daskaloyannis, M.E. Grypeos, C.G.
Koutroulos, S.E. Massen, and D.S. Saloupis, Phys. Lett. {\bf
121B} (1983) 91. 

\bibitem{lp93} G.A. Lalazissis and C.P. Panos, J. Phys. G: Nucl.
Part. Phys., {\bf 19} (1993) 283.
 
\bibitem{kra90} G.J. Kramer, Ph.D. Thesis, Amsterdam 1990
(unpublished). 

\bibitem{mou76} J. Mougey et al, Nucl. Phys. {\bf A262} (1976) 461.


\bibitem{lip85} S. Adachi, E. Lipparini, Nguyen Van Giai, Nucl.
Phys. {\bf A438} (1985) 1.

\bibitem{lp91} K. Takayanagi, E. Lipparini, Phys. Lett. {\bf
B261} (1991) 11.

\bibitem{neck92} D. Van Neck, M. Waroquier, V. Van Sluys and J.
Ryckebusch, Phys. Lett. {\bf B274} (1992) 143.



\end{thebibliography}
\end{document}